\documentstyle[myart,12pt]{article}
\normalbaselines
\font\tenbm=cmmib10
\font\sevenbm=cmmib7
\font\fivebm=cmmib5
\newfam\bmfam
\textfont\bmfam=\tenbm \scriptfont\bmfam=\sevenbm
\scriptscriptfont\bmfam=\fivebm
{\count0=\number\bmfam \multiply\count0 by "100
\def\defbgreek#1#2#3{{\count1=\count0 \advance\count1 by "#2#3
  \global\mathchardef#1=\count1 }}
\defbgreek\balpha  0B \defbgreek\brho       1A
\defbgreek\bbeta   0C \defbgreek\bsigma     1B
\defbgreek\bgamma  0D \defbgreek\btau       1C
\defbgreek\bdelta  0E \defbgreek\bupsilon   1D
\defbgreek\bepsilon0F \defbgreek\bphi       1E
\defbgreek\bzeta   10 \defbgreek\bchi       1F
\defbgreek\bmeta   11 \defbgreek\bpsi       20
\defbgreek\btheta  12 \defbgreek\bomega     21
\defbgreek\biota   13 \defbgreek\bvarepsilon22
\defbgreek\bkappa  14 \defbgreek\bvartheta  23
\defbgreek\blambda 15 \defbgreek\bvarpi     24
\defbgreek\bmu     16 \defbgreek\bvarrho    25
\defbgreek\bnu     17 \defbgreek\bvarsigma  26
\defbgreek\bxi     18 \defbgreek\bvarphi    27
\defbgreek\bpi     19}

\begin{document}

\title{Hamilton Variational Principle for Statistical Ensemble of Deterministic
Systems and its Application for Esemble of Stochastic Systems}
\author{Yuri A.Rylov}
\date{Institute for Problems in Mechanics, Russian Academy of Sciences,\\
101, bild.1 Vernadskii Ave., Moscow, 117526, Russia.\\
e-mail: rylov@ipmnet.ru}
\maketitle

\begin{abstract}
Hamilton variational principle for a special type of statistical ensemble of
deterministic dynamical systems is derived. This form of variational
principle allows one to describe the statistical ensemble in terms of wave
functions and provides a basis for a description of statistical ensemble of
stochastic systems. It is shown that sometimes such a statistical
description of stochastic particle motion appears to coincide with the
quantum description in terms of Schr\"odinger equation.
\end{abstract}

PACS\ numbers: \quad 03.40.Gc; 47.10.+g; 03.65.Bz

\newpage

\section{Introduction}

One derives the Hamilton variational principle, which appears to be an
effective mathematical tool of a statistical description. It allows one to
write dynamic equations for a statistical ensemble in terms of wave
functions and to show that the quantum mechanical description in terms of
the Schr\"odinger equation is a special case of statistical description.

Let a deterministic dynamic system ${\cal S}_{{\rm d}}$\footnote{%
The term ''dynamic system'' is used as a collective concept with respect to
terms ''nondeterministic (stochastic) dynamic system'' and ''deterministic
dynamic system''. Such an unusual teminology is used, because we need a
collective concept for concepts ''dynamic system' and ''stochastic system''.}
be described by the Lagrangian $L(t,{\bf x},d{\bf x}/dt),$ where ${\bf x}%
=\left\{ x^\alpha \right\} \;\;\alpha =1,2,...n,$ are generalized
coordinates and ${\bf v}=d{\bf x}/dt=\left\{ dx^\alpha /dt\right\}
\;\;\alpha =1,2,...n$ is generalized velocity. For brevity we shall speak
about a particle with a position ${\bf x}$ and velocity ${\bf v}$. A
statistical ensemble ${\cal E}_{{\rm gen}}\left[ {\cal S}_{{\rm d}}\right] $
(i.e. a set of many independent identical systems ${\cal S}_{{\rm d}}$) is a
deterministic dynamic system, whose state is described by the distribution
function $F\left( t,{\bf x},{\bf p}\right) $, where ${\bf p}=\left\{
\partial L/\partial (dx^\alpha /dt)\right\} ,\;\;\alpha =1,2,...n$ is the
generalized momentum of ${\cal S}_{{\rm d}}$. The state $F\left( t,{\bf x},%
{\bf p}\right) $ of the general ensemble ${\cal E}_{{\rm gen}}\left[ {\cal S}%
_{{\rm d}}\right] $ evolves according to the Liouville equation
\begin{equation}
{\cal E}_{{\rm gen}}\left[ {\cal S}_{{\rm d}}\right] :\qquad \frac{\partial F%
}{\partial t}+\frac \partial {\partial x^\alpha }\left( \frac{\partial H}{%
\partial p_\alpha }F\right) -\frac \partial {\partial p_\alpha }\left( \frac{%
\partial H}{\partial x^\alpha }F\right) =0  \label{a1.1}
\end{equation}
where $H=H\left( t,{\bf x},{\bf p}\right) ={\bf vp}-L$ is the Hamilton
function for the dynamic system ${\cal S}_{{\rm d}}$. Summation is produced
over repeating Greek indices $(1-n)$. Dynamic systems ${\cal S}_{{\rm d}}$,
constituting the statistical ensemble ${\cal E}_{{\rm gen}}\left[ {\cal S}_{%
{\rm d}}\right] $, are called elements of this ensemble. Dynamic equation (%
\ref{a1.1}) for the statistical ensemble ${\cal E}_{{\rm gen}}\left[ {\cal S}%
_{{\rm d}}\right] $ is obtained formally from the dynamic equations for
elements ${\cal S}_{{\rm d}}$ of the statistical ensemble ${\cal E}_{{\rm gen%
}}\left[ {\cal S}_{{\rm d}}\right] $.

We are interested in a special type ${\cal E}_{{\rm p}}\left[ {\cal S}_{{\rm %
d}}\right] $ of the general ensemble ${\cal E}_{{\rm gen}}\left[ {\cal S}_{%
{\rm d}}\right] $ which will be referred to as a pure statistical ensemble,
because its state may be described by the wave function\footnote{%
In quantum mechanics a pure state of a quantum system  is called a such one
which can be described by means of wave function (not by the dense matrix)
\cite{N32}.}. By definition a pure statistical ensemble ${\cal E}_{{\rm gen}%
}\left[ {\cal S}_{{\rm d}}\right] $ of dynamical systems ${\cal S}_{{\rm d}}$
is such an ensemble, whose state $F_{{\rm p}}\left( t,{\bf x},{\bf p}\right)
$ may be represented in the form
\begin{equation}
{\cal E}_{{\rm p}}\left[ {\cal S}_{{\rm d}}\right] :\qquad F_{{\rm p}}\left(
t,{\bf x},{\bf p}\right) =\rho \left( t,{\bf x}\right) \delta \left( {\bf p}-%
{\bf P}\left( t,{\bf x}\right) \right)  \label{a1.2}
\end{equation}
where $\rho \left( t,{\bf x}\right) $ and ${\bf P}\left( t,{\bf x}\right)
=\left\{ P_\alpha \left( t,{\bf x}\right) \right\} $, $\alpha =1,2,...n$ are
functions of only time $t$ and generalized coordinates ${\bf x}$. In other
words, the pure ensemble ${\cal E}_{{\rm p}}\left[ {\cal S}_{{\rm d}}\right]
$ is a dynamic system, considered in the configuration space of coordinates $%
{\bf x}$.

Dynamic equations for the pure ensemble ${\cal E}_{{\rm p}}\left[ {\cal S}_{%
{\rm d}}\right] $ have the form
\begin{equation}
\frac \partial {\partial t}\rho +\frac \partial {\partial x^\alpha }\left(
\rho \frac{\partial H}{\partial P_\alpha }\left( t,{\bf x},{\bf P}\right)
\right) =0  \label{a1.2a}
\end{equation}
\begin{equation}
\frac \partial {\partial t}\left( \rho P_\beta \right) +\frac \partial
{\partial x^\alpha }\left( \rho P_\beta \frac{\partial H}{\partial P_\alpha }%
\left( t,{\bf x},{\bf P}\right) \right) +\rho \frac{\partial H}{\partial
x^\beta }\left( t,{\bf x},{\bf P}\right) =0,\qquad \beta =1,2,...n
\label{a1.2b}
\end{equation}
They are obtained after substituting (\ref{a1.2}) into (\ref{a1.1}) and
integration with respect to ${\bf p}\ $with the multipliers $1,p_\beta
,\;\;\beta =1,2,...n$ respectively. Interpreting $\rho $ as a particle
density and ${\bf v}=\partial H/\partial {\bf p}$ as a generalized velocity,
the equation (\ref{a1.2a}) is regarded as a continuity equation. Then (\ref
{a1.2b}) may be interpreted as generalized Euler equations for some ideal
fluid without pressure.

Use of pure statistical ensembles is convenient in the case, when elements
of the statistical ensemble are nondeterministic (stochastic) dynamic
systems ${\cal S}_{{\rm st}},$ for which there are no dynamic equations.
Then the following procedure is used. The general statistical ensemble $%
{\cal E}_{{\rm gen}}\left[ {\cal S}_{{\rm st}}\right] $ is considered to be
consisting of elements ${\cal E}_{{\rm p}}\left[ {\cal S}_{{\rm st}}\right] $%
, which consist in turn of elements ${\cal S}_{{\rm st}}$, i.e.
\begin{equation}
{\cal E}_{{\rm gen}}\left[ {\cal S}_{{\rm st}}\right] ={\cal E}_{{\rm gen}%
}\left[ {\cal E}_{{\rm p}}\left[ {\cal S}_{{\rm st}}\right] \right]
\label{a1.3}
\end{equation}

There are no dynamic equations for dynamic system ${\cal S}_{{\rm st}}$, but
there are dynamic equations for a statistical ensemble ${\cal E}_{{\rm p}%
}\left[ {\cal S}_{{\rm st}}\right] $. Derivation of dynamic equations for $%
{\cal E}_{{\rm p}}\left[ {\cal S}_{{\rm st}}\right] $ is rather complicated
informal procedure, which needs a special consideration for any kind of
stochastic system ${\cal S}_{{\rm st}}$. But if dynamic equations for
dynamic systems ${\cal E}_{{\rm p}}\left[ {\cal S}_{{\rm st}}\right] $ have
been derived, dynamic equations for the statistical ensemble ${\cal E}_{{\rm %
gen}}\left[ {\cal E}_{{\rm p}}\left[ {\cal S}_{{\rm st}}\right] \right] $ of
deterministic dynamic systems ${\cal E}_{{\rm p}}\left[ {\cal S}_{{\rm st}%
}\right] $ are obtained as a result of a formal procedure. Thus, complicated
informal procedure of dynamic equations derivation for ${\cal E}_{{\rm gen}%
}\left[ {\cal S}_{{\rm st}}\right] $ is divided into two parts: (1)
derivation of dynamic equations for a more simple statistical ensemble $%
{\cal E}_{{\rm p}}\left[ {\cal S}_{{\rm st}}\right] $ by means of informal
procedure, (2) derivation of dynamic equations for ${\cal E}_{{\rm gen}%
}\left[ {\cal E}_{{\rm p}}\left[ {\cal S}_{{\rm st}}\right] \right] $ by
means of a formal procedure.

In this sense, a consideration of more simple pure statistical ensembles $%
{\cal E}_{{\rm p}}$ simplifies investigation of stochastic systems.

Let us add to the Euler equations the following equations
\begin{equation}
\frac{dx^\beta }{dt}=\frac{\partial H}{\partial P_\beta }\left( t,{\bf x},%
{\bf P}\right) ,\qquad \beta =1,2,...n  \label{a1.3c}
\end{equation}
describing motion of a particle in a given velocity field ${\bf v=}\partial
H/\partial {\bf P}$. These equations can be rewritten in the form, known in
hydrodynamics as Lin constraints \cite{L63}
\begin{equation}
\frac{\partial \xi _\beta }{\partial t}+\frac{\partial H}{\partial P_\alpha }%
\left( t,{\bf x},{\bf P}\right) \partial _\alpha \xi _\beta =0,\qquad \beta
=1,2,...n  \label{a1.4c}
\end{equation}
where ${\bf \xi }\left( t,{\bf x}\right) {\bf =}\left\{ \xi _\alpha \left( t,%
{\bf x}\right) \right\} ,\;\;\;\alpha =1,2,...n$ are $n$ independent
integrals of $n$ equations (\ref{a1.3c}).

The system of $2n+1$ equations (\ref{a1.2a}), (\ref{a1.2b}), (\ref{a1.4c})
forms a complete system of dynamic equations, describing evolution of
dynamic systems ${\cal S}_{{\rm d}}$, constituting the pure ensemble
${\cal E}_{{\rm p}}\left[ {\cal S}_{{\rm d}}\right] $.

One can show that the system of $2n+1$ equations (\ref{a1.2a}), (\ref{a1.2b}%
), (\ref{a1.4c}) reduces to the form of a system of $n+2$ equations
\begin{equation}
b_0[\partial _0\varphi +g^\alpha (\bxi )\partial _0\xi _\alpha ]+H\left(
{\bf x,P}\right) =0,\qquad \partial _k\equiv \frac \partial {\partial
x^k},\qquad k=0,1,...n  \label{a1.6c}
\end{equation}
\begin{equation}
\partial _0\rho +\partial _\alpha \left( \rho \frac{\partial H}{\partial
P_\alpha }\left( t,{\bf x},{\bf P}\right) \right) =0  \label{a1.2aa}
\end{equation}
\begin{equation}
\frac{\partial \xi _\beta }{\partial t}+\frac{\partial H}{\partial P_\alpha }%
\left( t,{\bf x},{\bf P}\right) \partial _\alpha \xi _\beta =0,\qquad \beta
=1,2,...n  \label{a1.4cc}
\end{equation}
where $\varphi $ is a new variable, and ${\bf P}$ is expressed now via $n$
arbitrary functions ${\bf g}\left( \bxi \right) =\left\{ g^\alpha \left( %
\bxi \right) \right\} ,\;\;\;\alpha =1,2,...n$ of argument $\bxi .$
\begin{equation}
P_\beta =b_0\left( \partial _\beta \varphi +g^\alpha \left( \bxi \right)
\right) \partial _\beta \xi _\alpha ,\qquad \beta =1,2,...n  \label{a1.5c}
\end{equation}
Practically it means that the system of $2n+1$ equations (\ref{a1.2aa}), (%
\ref{a1.2b}), (\ref{a1.4cc}) is integrated partially. Here $b_0$ is an
arbitrary constant, which can be incorporated in the variable $\varphi $ and
arbitrary functions ${\bf g}\left( \bxi \right) $.

The system of $n+2$ equations (\ref{a1.6c}), (\ref{a1.2aa}), (\ref{a1.4cc})
is remarkable in the relation that it can be described in terms of $\psi $
function (wave function). But it is more convenient one to transform these
equations, using the Hamilton variational principle.

We shall show that dynamic equations for the pure statistical ensemble $%
{\cal E}_{{\rm p}}\left[ {\cal S}_{{\rm d}}\right] $ of deterministic
dynamic systems ${\cal S}_{{\rm d}}$ are derived from the variational
principle with the action
\begin{equation}
{\cal E}_{{\rm p}}\left[ {\cal S}_{{\rm d}}\right] :\qquad {\cal A}_E[\rho
,\varphi ,\bxi ]=\int \rho \{-H\left( t,{\bf x,p}\right) -b_0[\partial
_0\varphi +g^\alpha (\bxi )\partial _0\xi _\alpha ]\}d^{n+1}x,  \label{a1.4}
\end{equation}
\begin{equation}
p_\beta =b_0\left[ \partial _\beta \varphi +g^\alpha (\bxi )\partial _\beta
\xi _\alpha \right] ,\qquad \beta =1,2,...,n,,\qquad \partial _i\equiv \frac
\partial {\partial x^i}  \label{a1.5}
\end{equation}
where $\rho ,\varphi ,\bxi $ are dependent variables, which are considered
to be functions of $x=\left\{ x^0,{\bf x}\right\} =\left\{ t,{\bf x}\right\}
$. $H\left( t,{\bf x,p}\right) $ is the Hamiltonian function of ${\cal S}_{%
{\rm d}}$. $b_0$ is an arbitrary constant and $g^\alpha (\bxi ),\;\;\alpha
=1,2,...,n$ are arbitrary functions of the argument $\bxi $. The dynamic
variables $\varphi ,\bxi $ are hydrodynamic (Clebsch) potentials. Clebsch
\cite{C57,C59} had introduced them for a description of incompressible
fluid. The variables $\varphi ,\bxi $ are called potentials, because
momentum ${\bf p}={\bf P}\left( t,{\bf x}\right) $ is expressed via
derivatives of potentials $\varphi ,\bxi $, as one can see from the
relations (\ref{a1.5}). The Hamiltonian $H\left( t,{\bf x,p}\right) $ is a
function which determines the form of the action (\ref{a1.4}), and the
variational principle based on the action (\ref{a1.4}) may be called the
Hamilton variational principle.

The dynamic system ${\cal E}_{{\rm p}}\left[ {\cal S}_{{\rm d}}\right] $ is
an ideal fluidlike continuous medium without pressure. Variational principle
for such a system in the Lagrangian coordinates $\bxi ,$ (where $\bxi $ are
considered to be independent variables) is derived very simple. But a
derivation of a variational principle for an ideal fluid in Euler
coordinates ${\bf x}$ was being a problem for many years \cite{S88}, because
one tried to obtain a variational principle only for the Euler equations,
i.e. for equations of the type (\ref{a1.2a}), (\ref{a1.2b}). The Euler
equations form a closed subsystem of the complete system of dynamic
equations. The variational principle generates only a complete system of
dynamic equations (but not their closed subsystem). When the Lin constraints
\cite{L63} were added to the Euler equations, the system of dynamic
equations became complete, and the variational principle was written for
different partial cases \cite
{S88,D49,H55,E60,C63,SW67,Br70,FS78,S82,B83,ZK97}. The variational principle
was written for an ideal fluid which can be considered to be a continuous
set ${\cal S}\left[ {\cal S}_{{\rm d}}\right] $ of nonrelativistic particles
${\cal S}_{{\rm d}}$, interacting between themselves via the pressure, which
is a function of a collective variable (particle density) $\rho $ and
temperature $T$. Hamiltonian function of ${\cal S}_{{\rm d}}$ has the form
\[
H\left( t,{\bf x,p}\right) =\frac{{\bf p}^2}{2m}+V\left( {\bf x}\right) ,
\]
The case, when the pressure is a function of $\rho ,$ ${\bf \nabla }\rho $
and $T$, was considered in the paper \cite{R99}. Now the case of arbitrary
Hamiltonian function of ${\cal S}_{{\rm d}}$ is considered. Parameters of
the Hamiltonian (not necessarily the pressure) are considered to be
functions of $\rho $ and its space-time derivatives.

In the second section the Hamilton variational principle is derived. In the
third section one shows how the Hamilton variational principle generates a
description in terms of $\psi $-function (wave function). In the fourth
section ''conservative'' stochastic systems are described on the basis of
the Hamilton variational principle

\section{Variational principle}

The action for the deterministic dynamic system ${\cal S}_{{\rm d}}$ has the
form
\begin{equation}
{\cal S}_{{\rm d}}:\qquad {\cal A}_L[{\bf x}]=\int L(t,{\bf x},d{\bf x}%
/dt)dt,  \label{a2.1}
\end{equation}
A set of many independent systems ${\cal S}_{{\rm d}}$ forms a deterministic
dynamic system (statistical ensemble) ${\cal E}_{{\rm p}}\left[ {\cal S}_{%
{\rm d}}\right] $, whose action is a sum of actions (\ref{a2.1}). Let us
label each system ${\cal S}_{{\rm d}}$ by parameters ${\bxi =}\left\{ \xi
_1,\xi _2,...\xi _n\right\} $, where $n$ is the number of generalized
coordinates. Usually the variables $\bxi $ are referred to as Lagrangian
coordinates or generalized Lagrangian coordinates. One obtains for the
action of the dynamic system ${\cal E}_{{\rm p}}\left[ {\cal S}_{{\rm d}%
}\right] $%
\begin{equation}
{\cal E}_{{\rm p}}\left[ {\cal S}_{{\rm d}}\right] :\qquad {\cal A}_L[{\bf x}%
]=\int \rho _0\left( \bxi \right) L(t,{\bf x},d{\bf x}/dt){\rm d}t{\rm d}%
\bxi ,\qquad {\rm d}\bxi =\prod\limits_{\alpha =1}^{\alpha =n}{\rm d}\xi
_\alpha ,  \label{a2.2}
\end{equation}
where $\rho _0\left( \bxi \right) $ is some nonnegative weight function. For
simplicity we set $\rho _0=1$, because in reality a use of the weight
function is unessential. One can see this for a special case of the
continuous dynamic system ${\cal S}\left[ {\cal S}_{{\rm d}}\right] $,
considered in paper \cite{R99}, where the weight function was used. For
further calculations it is useful one to introduce a fictitious Lagrangian
time coordinate $\xi _0=\xi _0\left( t,\bxi \right) ,$ and rewrite the
action (\ref{a2.2}) in the form
\begin{equation}
{\cal E}_{{\rm p}}\left[ {\cal S}_{{\rm d}}\right] :\qquad {\cal A}_L[{\bf x}%
]=\int L(x,\frac{{\bf \dot x}}{\dot x^0})\dot x^0{\rm d}^{n+1}\xi ,\qquad
{\rm d}^{n+1}\xi =\prod\limits_{k=0}^{k=n}{\rm d}\xi _k,  \label{a2.3}
\end{equation}
where $x=\left\{ x^0,{\bf x}\right\} =\left\{ t,{\bf x}\right\} =\{ x^k\}
,\;\;k=0,1,...n,\;\;\xi =\left\{ \xi _0,\bxi \right\} =\left\{ \xi
_k\right\} ,\;\;k=0,1,...n$, and the point means differentiation with
respect to $\xi _0$%
\begin{equation}
{\bf \dot x}\equiv \frac{d{\bf x}}{d\xi _0},\qquad \dot x^0\equiv \frac{dx^0%
}{d\xi _0}  \label{a2.4}
\end{equation}

Now the action (\ref{a2.3}) is considered to be a functional of $n$
dependent variables ${\bf x}.$ The variable $t=x^0$ is some fixed function
of $\xi $. The form of this function is unessential.

Let us consider the Lagrangian coordinates $\bxi $ to be dependent
variables, which are functions of independent variables -- coordinates
$x=\left\{ t,{\bf x%
}\right\} =\left\{ x^0,{\bf x}\right\} =\left\{ x^i\right\} ,\;\;i=0,1,...n$
and transform the action to the form, where Eulerian coordinates $x$ are
independent variables. Let us introduce designations
\begin{equation}
J=\frac{\partial \left( \xi _0,\xi _1,...\xi _n\right) }{\partial \left(
x^0,x^1,...x^n\right) }=\det ||\xi _{i,k}||,\qquad \xi _{i,k}\equiv \frac{%
\partial \xi _i}{\partial x^k}\equiv \partial _k\xi _i\qquad i,k=0,1,...n
\label{a2.5}
\end{equation}
\begin{equation}
j^i=\frac{\partial J}{\partial \xi _{0,i}}=\frac{\partial \left( x^i,\xi
_1,\xi _2,...\xi _n\right) }{\partial \left( x^0,x^1,...x^n\right) }=J\dot
x^i,\qquad i=0,1,...n,\qquad \rho =j^0=\frac{\partial J}{\partial \xi _{0,0}}
\label{a2.6}
\end{equation}
and relations
\begin{equation}
{\rm d}\xi _0{\rm d}\bxi =J{\rm d}^{n+1}x\equiv J\prod\limits_{i=0}^{i=n}%
{\rm d}x^i,\qquad \frac{dx^\alpha }{dt}=\frac{\partial \left( x^\alpha ,\xi
_1,...\xi _n\right) }{\partial \left( t,\xi _1,...\xi _n\right) }=\frac{%
j^\alpha }{j^0},\qquad \alpha =1,2,...n.  \label{a2.7}
\end{equation}
The action (\ref{a2.3}) transforms to the form
\begin{equation}
{\cal E}_{{\rm p}}\left[ {\cal S}_{{\rm d}}\right] :\qquad {\cal A}_E[\xi
]=\int {\cal L}\left( x,j\right) {\rm d}^{n+1}x,  \label{a2.8}
\end{equation}
\[
{\cal L}\left( x,j\right) =L(x^0,{\bf x},\frac{{\bf j}}{j^0})j^0
\]
where $j=\left\{ j^0,{\bf j}\right\} $ is a function of $\xi _{i,k}$,
defined by the relations (\ref{a2.6}). The action (\ref{a2.8}) is considered
to be a functional of $n+1$ dependent variables $\xi $. In fact the variable
$\xi _0$ is fictitious, and variation with respect to $\xi _0$ leads to an
identity.

Let us introduce new variables $j=\left\{ j^0,{\bf j}\right\} =\left\{
j^k\right\} ,\;\;k=0,1,...n$, defined by the relation (\ref{a2.6}) by means
of the Lagrangian multipliers $p=\left\{ p_k\right\} ,\;\;k=0,1,...n$.
\begin{equation}
{\cal E}_{{\rm p}}\left[ {\cal S}_{{\rm d}}\right] :\qquad {\cal A}_E[\xi
,p,j]=\int \left( {\cal L}\left( x,j\right) -p_i\left( j^i-\frac{\partial J}{%
\partial \xi _{0,i}}\right) \right) {\rm d}^{n+1}x,  \label{a2.9}
\end{equation}
Here and further a summation over repeated Latin super- and subindexes is
made from $0$ to $n$.

There are two ways for derivation of dynamic equations. The first way: the
action (\ref{a2.9}) is varied with respect to variables $j,p,{\bxi
}$, and after elimination of Lagrangian coordinates $\xi $ the Eulerian
equations (\ref{a1.2a}), (\ref{a1.2b}) appear. We are interested in another
way, when dynamic equations $\delta {\cal A}_E/\delta \xi _i=0$ are
integrated in the form (\ref{a1.5}) and after elimination of variables $p,j$%
, one obtains dynamic equations of the Hamilton-Jacobi type for hydrodynamic
potentials $\xi $.

Variation of the action (\ref{a2.9}) with respect to $\xi _k$ leads to
dynamic equations
\begin{equation}
\delta \xi _k:\qquad -\partial _l\left( p_i\frac{\partial ^2J}{\partial \xi
_{0,i}\partial \xi _{k,l}}\right) =0  \label{a2.10}
\end{equation}
Being linear with respect to $p,$ equations (\ref{a2.10}) can be solved in
the form
\begin{equation}
p_k=b_0\left( \partial _k\varphi +g^\alpha \left( \bxi \right) \right)
\partial _k\xi _\alpha ,\qquad k=0,1,...n  \label{a2.14}
\end{equation}
where $b_0$ is an arbitrary constant, $g^\alpha \left( \bxi \right)
,\;\;\alpha =1,2,...n$ are arbitrary integration functions of the variables $%
{\bxi .}$ After integration the fictitious variable $\xi _0$ stops to be
fictitious, and $\varphi $ is a new dependent variables which appears
instead of $\xi _0$. Using Jacobian technique \cite{R99}, one can verify by
means of a direct substitution of (\ref{a2.14}) in (\ref{a2.10}) that (\ref
{a2.14}) is a solution of (\ref{a2.10}). In particular, one needs the
following identities
\begin{equation}
\partial _k\frac{\partial ^2J}{\partial \xi _{i,k}\partial \xi _{s,l}}\equiv
0,\qquad i,s,l=0,1,...n.  \label{a2.14a}
\end{equation}
\begin{equation}
\frac{\partial ^2J}{\partial \xi _{i,k}\partial \xi _{s,l}}\equiv \frac
1J\left( \frac{\partial J}{\partial \xi _{i,k}}\frac{\partial J}{\partial
\xi _{s,l}}-\frac{\partial J}{\partial \xi _{i,l}}\frac{\partial J}{\partial
\xi _{s,k}}\right) ,\qquad i,k,l,s=0,1,...n,  \label{a2.15}
\end{equation}

\begin{equation}
\xi _{l,k}\frac{\partial J}{\partial \xi _{s,k}}\equiv \delta _l^sJ,\qquad
\xi _{k,l}\frac{\partial J}{\partial \xi _{k,s}}\equiv \delta _l^sJ,\qquad
l,s=0,1,...n,  \label{a2.16}
\end{equation}
\begin{equation}
\partial _k\frac{\partial J}{\partial \xi _{i,k}}\equiv \frac{\partial ^2J}{%
\partial \xi _{i,k}\partial \xi _{s,l}}\partial _k\partial _l\xi _s\equiv
0,\qquad i=0,1,...n.  \label{a2.17}
\end{equation}
Eliminating $p_k$ from relations (\ref{a2.14}) and (\ref{a2.9}), one obtains
the action
\begin{equation}
{\cal A}_E[j,\varphi ,\bxi ]=\int \{{\cal L}(x,j)-b_0j^i[\partial _i\varphi
+g^\alpha (\bxi )\partial _i\xi _\alpha ]\}{\rm d}^{n+1}x,  \label{a2.18}
\end{equation}

Let us introduce new variables
\[
{\bf v=j/}j^0,\qquad \rho =j^0
\]
Then the action (\ref{a2.18}) transforms to the form
\[
{\cal A}_E[\rho ,{\bf v},\varphi ,\bxi ]=\int \rho \{L(x,{\bf v}%
)-b_0[\partial _0\varphi +g^\alpha (\bxi )\partial _0\xi _\alpha ]
\]
\begin{equation}
-b_0v^\beta [\partial _\beta \varphi +g^\alpha (\bxi )\partial _\beta \xi
_\alpha ]\}{\rm d}^{n+1}x,  \label{a2.19}
\end{equation}
Let us now eliminate the variables ${\bf v}.$ Varying the action (\ref{a2.19}%
) with respect to $v^\beta ,$ one obtains
\begin{equation}
\frac{\delta {\cal A}_E}{\delta v^\beta }=\rho \left( \frac{\partial L(x,%
{\bf v})}{\partial v^\beta }-b_0[\partial _\beta \varphi +g^\alpha ({\bxi }%
)\partial _\beta \xi _\alpha ]\right) =0,\qquad \beta =1,2,...n
\label{a2.20}
\end{equation}
Let us use designations (\ref{a2.14}). The dynamic equations (\ref{a2.20})
take the form
\begin{equation}
\rho \left( \frac{\partial L(x,{\bf v})}{\partial v^\beta }-p_\beta \right)
=0,\qquad \beta =1,2,...n  \label{a2.22}
\end{equation}

If the dynamic system ${\cal S}_{{\rm d}}$ is Hamiltonian, $n$ equations (%
\ref{a2.22}), can be resolved with respect to ${\bf v}$ in the form $v^\beta
=\partial H/\partial p_\beta ,\;\;\;\beta =1,2,...n$, where
\begin{equation}
H=H\left( t,{\bf x,p}\right) =p_\alpha v^\alpha -L(x,{\bf v})  \label{a2.23}
\end{equation}
is the Hamilton function for the dynamic system ${\cal S}_{{\rm d}}$.

Substituting ${\bf v}$ in the action (\ref{a2.19}), one obtains
\begin{equation}
{\cal A}_E[\rho ,\varphi ,\bxi ]=\int \rho \{-H\left( t,{\bf x,p}\right)
-b_0[\partial _0\varphi +g^\alpha (\bxi )\partial _0\xi _\alpha ]\}{\rm d}%
^{n+1}x,  \label{a2.24}
\end{equation}
where ${\bf p}$ is determined by the relation (\ref{a2.14}).

Dynamic equations have the form
\begin{equation}
\delta \rho :\qquad H\left( {\bf x,p}\right) +b_0[\partial _0\varphi
+g^\alpha (\bxi )\partial _0\xi _\alpha ]=0  \label{a2.25}
\end{equation}
\begin{equation}
\delta \varphi :\qquad b_0\left( \partial _0\rho +\partial _\beta \left(
\rho \frac{\partial H}{\partial p_\beta }\right) \right) =0  \label{a2.26}
\end{equation}
\begin{equation}
\delta \xi _\alpha :\qquad \Omega ^{\beta ,\alpha }\rho \left( \xi _{\beta
,0}+\frac{\partial H}{\partial p_\gamma }\xi _{\beta ,\gamma }\right)
=0,\qquad \Omega ^{\alpha ,\beta }\equiv \frac{\partial g^\alpha \left( \bxi %
\right) }{\partial \xi _\beta }-\frac{\partial g^\beta \left( \bxi \right) }{%
\partial \xi _\alpha }  \label{a2.27}
\end{equation}
where ${\bf p}$ is determined by the relation (\ref{a2.14}).

Let us consider a special case, when
\begin{equation}
\Omega ^{\alpha ,\beta }\equiv b_0\left( \frac{\partial g^\alpha \left( {\bf %
\xi }\right) }{\partial \xi _\beta }-\frac{\partial g^\beta \left( \bxi %
\right) }{\partial \xi _\alpha }\right) \equiv 0  \label{a2.28}
\end{equation}
Then
\[
g^\alpha \left( \bxi \right) =\frac{\partial \phi \left( \bxi \right) }{%
\partial \xi _\alpha },\qquad \alpha =1,2,...n,
\]
equations (\ref{a2.27}) are satisfied identically, and ${\bf p}$ is a
gradient
\begin{equation}
p_\beta =b_0\partial _\beta \left( \varphi +\phi \left( \bxi \right) \right)
,\qquad \beta =1,2,...n.  \label{a2.28a}
\end{equation}
Equation (\ref{a2.25}) turns to the Hamilton-Jacobi equation for the
variable $\Phi =b_0\left( \varphi +\phi \right) $%
\begin{equation}
\partial _0\Phi +H\left( {\bf x,p}\right) =0,\qquad {\bf p}={\bf \nabla }\Phi
\label{a2.29}
\end{equation}
In this case the dynamic system ${\cal E}_{{\rm p}}\left[ {\cal S}_{{\rm d}%
}\right] $, considered to be a generalized fluid, has an irrotational vector
field of momentum ${\bf p}$ (irrotational flow).

The case, when condition (\ref{a2.28}) is not satisfied, the momentum vector
field ${\bf p}$ is rotational (rotational flow). In this case equations (\ref
{a2.27}) may be considered to be some kind of generalization of the
Hamilton-Jacobi equation. For a real particle whose Hamiltonian has the form
\begin{equation}
H\left( {\bf x},{\bf p}\right) =\frac{{\bf p}^2}{2m}+V\left( {\bf x}\right)
\label{a2.30}
\end{equation}
the velocity ${\bf v}={\bf p}/m$, and vector fields of ${\bf p}$ and ${\bf v}
$ are simultaneously both rotational or irrotational. For a more general
form of Hamiltonian the vector field of ${\bf p}$ may be irrotational,
whereas the vector field of ${\bf v}$ is rotational, or vice versa.

Note that the action (\ref{a2.24}) contains information on initial values of
momenta ${\bf p}$ of dynamic system ${\cal E}_{{\rm p}}\left[ {\cal S}_{{\rm %
d}}\right] $. For instance, let us use standard initial conditions for
variables $\varphi ,\bxi $%
\begin{equation}
\bxi \left( 0,{\bf x}\right) ={\bf x,\qquad }\varphi \left( 0,{\bf x}\right)
=0,  \label{a2.31}
\end{equation}
i.e. elements ${\cal S}_{{\rm d}}$ of ${\cal E}_{{\rm p}}\left[ {\cal S}_{%
{\rm d}}\right] $ are labeled by their coordinates ${\bf x}$ at the moment $%
t $ $=0$. Then the integration functions $g^\alpha \left( \bxi \right) $ are
determined from the relation
\begin{equation}
{\bf p}\left( 0,{\bf x}\right) =b_0g^\alpha \left( \bxi \right) {\bf \nabla
\xi }_\alpha =b_0{\bf g}\left( \bxi \right) ,\qquad {\bf g}\left( \bxi %
\right) =\left\{ g^\alpha \left( \bxi \right) \right\} ,\qquad \alpha
=1,2,...n  \label{a2.32}
\end{equation}
Thus, one can consider that the action (\ref{a2.24}) contains complete
information on motion of element ${\cal S}_{{\rm d}}$ of ${\cal E}_{{\rm p}%
}\left[ {\cal S}_{{\rm d}}\right] $. Vice versa, if one knows how any
element ${\cal S}_{{\rm d}}$ moves, one knows all about motion of the
ensemble elements ${\cal S}_{{\rm d}}$ except for density of elements ${\cal %
S}_{{\rm d}}$ of ${\cal E}_{{\rm p}}\left[ {\cal S}_{{\rm d}}\right] ,$
described by the quantity $\rho $. The variable $\rho $ is a collective
variable, containing information on density of elements ${\cal S}_{{\rm d}}$
of the statistical ensemble ${\cal E}_{{\rm p}}\left[ {\cal S}_{{\rm d}%
}\right] $. Information on initial value of $\rho $ is to be given
additionally. Dynamic equations (\ref{a2.25}), (\ref{a2.27}), describing
motion of single elements ${\cal S}_{{\rm d}}$ of the statistical ensemble $%
{\cal E}_{{\rm p}}\left[ {\cal S}_{{\rm d}}\right] $, do not depend on the
quantity $\rho $. In general, evolution of the statistical ensemble is
insensitive to the number $N$ of its elements. This property of the
statistical ensemble is described by the relation
\begin{equation}
{\cal A}_E[a\rho ,{\bf v},\varphi ,\bxi ]=a{\cal A}_E[\rho ,{\bf v},\varphi ,%
\bxi ],\qquad a=\mbox{const}>0  \label{a2.33}
\end{equation}
The property (\ref{a2.33}) is valid for any statistical ensemble ${\cal E}_{%
{\rm p}}\left[ {\cal S}_{{\rm d}}\right] $ and ${\cal E}_{{\rm p}}\left[
{\cal S}_{{\rm st}}\right] $.

\section{Description in terms of $\psi $-function}

Let us introduce $k$-component complex function $\psi =\{\psi _\alpha
\},\;\;\alpha =1,2,\ldots k$, defining it by the relations

\[
\psi _\alpha =\sqrt{\rho }e^{i\varphi }u_\alpha (\bxi ),\qquad \psi _\alpha
^{*}=\sqrt{\rho }e^{-i\varphi }u_\alpha ^{*}(\bxi ),\qquad \alpha
=1,2,\ldots k\label{s5.4}
\]
\[
\psi ^{*}\psi \equiv \sum_{\alpha =1}^k\psi _\alpha ^{*}\psi _\alpha
\]
where (*) means the complex conjugate, $u_\alpha (\bxi )$, $\;\alpha
=1,2,\ldots k$ are functions of only variables $\bxi $. They satisfy the
relations
\begin{equation}
-\frac i2\sum_{\alpha =1}^k(u_\alpha ^{*}\frac{\partial u_\alpha }{\partial
\xi _\beta }-\frac{\partial u_\alpha ^{*}}{\partial \xi _\beta }u_\alpha
)=g^\beta (\bxi ),\qquad \beta =1,2,...n,\qquad \sum_{\alpha =1}^ku_\alpha
^{*}u_\alpha =1  \label{s5.5}
\end{equation}
$k$ is such a natural number that equations (\ref{s5.5}) admit a solution.
In general $k$ may depend on the form of the arbitrary integration functions
${\bf g}=\{g^\beta (\bxi )\}$,\ $\beta =1,2,...n.$

It is easy to verify that
\begin{equation}
\rho =\psi ^{*}\psi ,\qquad p_l(\varphi ,\xi \bxi )=-\frac{ib_0}{2\psi
^{*}\psi }(\psi ^{*}\partial _l\psi -\partial _l\psi ^{*}\cdot \psi ),\qquad
l=0,1,...n  \label{s5.6}
\end{equation}
The variational problem with the action (\ref{a2.24}) appears to be
equivalent to the variational problem with the action functional
\begin{equation}
A[\psi ,\psi ^{*}]=\int \{\frac{ib_0}2(\psi ^{*}\partial _0\psi -\partial
_0\psi ^{*}\cdot \psi )-H\left( x,-\frac{ib_0}{2\psi ^{*}\psi }(\psi ^{*}%
{\bf \nabla }\psi -{\bf \nabla }\psi ^{*}\cdot \psi )\right) \psi ^{*}\psi \}%
{\rm d}^{n+1}x  \label{s5.8}
\end{equation}
where ${\bf \nabla }=\left\{ \partial _\alpha \right\} ,\;\;\alpha =1,2,...n$

Note that the function $\psi $ considered to be a function of independent
variables $\{t,{\bf x}\}$ is very indefinite in the sense that the same
state $\left\{ \rho \left( t,{\bf x}\right) ,{\bf P}\left( t,{\bf x}\right)
\right\} $ of the statistical ensemble ${\cal E}_{{\rm p}}\left[ {\cal S}_{%
{\rm d}}\right] $ may be described by different $\psi $-functions. There are
two reasons for such an indefiniteness. First, the functions $u_\alpha (\bxi %
)$ are not determined uniquely by differential equations (\ref{s5.5}).
Second, their arguments $\bxi $ as functions of $x$ are determined only to
within the relabeling transformation
\begin{equation}
\xi _\alpha \to \tilde \xi _\alpha =\tilde \xi _\alpha (\bxi ),\qquad D=\det
\parallel \partial \tilde \xi _\alpha /\partial \xi _\beta \parallel
=1,\qquad \alpha ,\beta =1,2,...n  \label{d1.16}
\end{equation}

Description of the statistical ensemble ${\cal E}_{{\rm p}}\left[ {\cal S}_{%
{\rm d}}\right] $ in terms of the function $\psi $ is more indefinite, than
the description in terms of the hydrodynamic potentials $\bxi $. Information
about initial and boundary conditions, contained in functions ${\bf g}(\bxi %
),$ is lost at the description in terms of the $\psi $-function.

Dynamic equations have the form
\begin{equation}
\delta \psi _\beta ^{*}:\qquad \left[ ib_0\partial _0-H+\frac{\partial H}{%
\partial p_\alpha }p_\alpha +\frac{ib_0}2\left( \frac{\partial H}{\partial
p_\alpha }{\bf \nabla +\nabla }\frac{\partial H}{\partial p_\alpha }\right)
\right] \psi _\beta =0,\qquad \beta =1,2,...k  \label{s5.7}
\end{equation}
\begin{equation}
\delta \psi _\beta :\qquad \left[ -ib_0\partial _0-H+\frac{\partial H}{%
\partial p_\alpha }p_\alpha -\frac{ib_0}2\left( \frac{\partial H}{\partial
p_\alpha }{\bf \nabla +\nabla }\frac{\partial H}{\partial p_\alpha }\right)
\right] \psi _\beta ^{*}=0,\qquad \beta =1,2,...k  \label{s5.7a}
\end{equation}
where $H=H\left( x,{\bf p}\right) $ and $\frac{\partial H}{\partial p_\alpha
}$ are considered to be operators of multiplication by these quantities, and
one has to substitute the expression (\ref{s5.6}) instead of ${\bf p}$
before action of operator ${\bf \nabla }$. In general, dynamic equations (%
\ref{s5.7}) (\ref{s5.7a}) are not linear with respect to $\psi $-function,
although they may be linear in some cases. In these interesting cases the
dynamic equations may be solved rather simply.

The number $k$ of the $\psi $-function components in the actions (\ref{s5.8}%
) is arbitrary. A formal variation of the action with respect to $\psi
_\alpha $ and $\psi _\alpha ^{*},\quad \alpha =1,2,\ldots k$ leads to $2k$
real dynamic equations, but not all of them are independent. There are such
combinations of variations $\delta \psi _\alpha $, $\delta \psi _\alpha ^{*}$%
, $\alpha =1,2,\ldots k$ which do not change expressions (\ref{s5.6}). Such
combinations of variations $\delta \psi _\alpha $, $\delta \psi _\alpha ^{*}$%
, $\alpha =1,2,\ldots k$ do not change the action (\ref{s5.8}), and
corresponding combinations of dynamic equations $\delta {\cal A}/\delta \psi
_\alpha =0$, $\delta {\cal A}/\delta \psi _\alpha ^{*}=0$ are identities
that associates with a correlation between dynamic equations. Thus,
increasing the number $k$, one increases the number of dynamic equations,
but the number of independent dynamic equations remains the same. The number
$k$ is restricted underside by the condition that equations (\ref{s5.5})
have a solution. In other words, the minimal number $k_m$ of the $\psi $%
-function components depends on the form of functions ${\bf g}(\bxi )$, i.e.
on initial conditions. This number $k_m$ associates with the kinematic spin
( $k$-spin) $s=2k_m+1$ of the ensemble state \cite{R99}.

The $\psi $-function and $k$-spin remind quantum mechanical wave function
and spin of a particle respectively. The $\psi $-function coincides with the
wave function, if the dynamic equations (\ref{s5.7}), (\ref{s5.7a}) become
linear. It appears to be possible for statistical ensemble ${\cal E}_{{\rm p}%
}\left[ {\cal S}_{{\rm st}}\right] $ of stochastic systems ${\cal S}_{{\rm st%
}}$. In this case the $k$-spin associates with the spin of the described
particle, but the $k$-spin remains to be a property of the statistical
ensemble ${\cal E}_{{\rm p}}\left[ {\cal S}_{{\rm st}}\right] $ ( i.e. a
collective property), whereas in quantum mechanics the spin is considered to
be a property of an individual particle.

\section{Pure Statistical Ensemble of stochastic particles}

Let us consider statistical ensemble ${\cal E}_{{\rm p}}\left[ {\cal S}_{%
{\rm st}}\right] $ of stochastic particles ${\cal S}_{{\rm st}}$. There are
no dynamic equations for ${\cal S}_{{\rm st}}$, and one cannot derive
dynamic equations for ${\cal E}_{{\rm p}}\left[ {\cal S}_{{\rm st}}\right] $
from dynamic equations for ${\cal S}_{{\rm st}}$. But we believe that
dynamic equations for ${\cal E}_{{\rm p}}\left[ {\cal S}_{{\rm st}}\right] $
exist, because experiments with statistical ensembles of stochastic
particles ${\cal S}_{{\rm st}}$ are reproducible. Let us consider motion of $%
{\cal S}_{{\rm st}}$ as a result of interaction between some deterministic
particle ${\cal S}_{{\rm d}}$ and some stochastic agent, which makes motion
of ${\cal S}_{{\rm d}}$ to be stochastic. To derive dynamic equations for $%
{\cal E}_{{\rm p}}\left[ {\cal S}_{{\rm st}}\right] $, some supposition on
properties of this stochastic agent are to be made, because one cannot
derive dynamic equations for ${\cal E}_{{\rm p}}\left[ {\cal S}_{{\rm st}%
}\right] $ from nothing. If ${\cal S}_{{\rm st}}$ is a Brownian particle,
moving in a gas, one supposes that the Brownian particle collides with gas
molecules, and these collisions make the Brownian particle motion to be
stochastic. One supposes that these collisions are accidental and
independent. As a result a motion of the Brownian particle may be considered
to be a Markovian process. Dynamic system ${\cal E}_{{\rm p}}\left[ {\cal S}%
_{{\rm st}}\right] $ appears to be dissipative in this case, and there is no
variational principle for it.

However, there is another kind of stochastic agent which remains the
statistical ensemble ${\cal E}_{{\rm p}}\left[ {\cal S}_{{\rm st}}\right] $
to be a conservative deterministic dynamic system. Dynamic equations of this
system can be derived from the Hamilton variational principle formulated in
a proper way. This stochastic agent is known as a quantum stochasticity,
which is an origin of quantum effects, described by quantum mechanics.
Stochastic systems ${\cal S}_{{\rm st}}$, associated with such a kind of
stochasticity will be referred to as conservative stochastic systems.

Let us consider some microparticle, for instance, electron. An individual
electron is a stochastic system ${\cal S}_{{\rm st}}$, because experiments
with a single electron are irreproducible. For instance, in a diffraction
experiment a single electron, flying through a narrow hole in a diaphragm,
hits a new point of a screen any time. But distribution of many independent
electrons over the screen surface is reproducible. It means, that the
statistical ensemble ${\cal E}_{{\rm p}}\left[ {\cal S}_{{\rm st}}\right] $
of many ${\cal S}_{{\rm st}}$ is a deterministic dynamic system ${\cal S}_{%
{\rm S}}$. This deterministic dynamic system ${\cal S}_{{\rm S}}$ is
described by the action
\begin{equation}
{\cal A}_{{\rm S}}\left[ \psi ^{*},\psi \right] =\int \{\frac{i\hbar }2(\psi
^{*}\partial _0\psi -\partial _0\psi ^{*}\cdot \psi )-\frac{\hbar ^2}{2m}%
{\bf \nabla }\psi ^{*}\cdot {\bf \nabla }\psi \}{\rm d}^4x  \label{c3.1}
\end{equation}
where $\psi $ is so called wave function. Dynamic equation, generated by the
action (\ref{c3.1})
\begin{equation}
i\hbar \partial _0\psi +\frac{\hbar ^2}{2m}{\bf \nabla }^2\psi =0
\label{c3.2}
\end{equation}
is known as the Schr\"odinger equation.

Why an electron is described in terms of wave function? What is the wave
function? Principles of quantum mechanics answer these questions. The wave
function is a fundamental mathematical object (a point in the Hilbert space)
which evolves according to linear dynamic equation. All relations of quantum
mechanics are obtained from corresponding relations of classical ones,
replacing position ${\bf x}$ $\rightarrow $ operator of multiplication by $%
{\bf x}$ and momentum ${\bf p\rightarrow -i\hbar \nabla }$. Quantum
mechanics is an axiomatic theory, and quantum principles work very well in
the non-relativistic case. Attempts of extension of the quantum mechanic
principles to the relativistic case met problems. Investigators were forced
to introduce corrections and new principles. All this leads to a suspicion
that quantum mechanical principles are conceptually non-relativistic, and
principles of quantum mechanics cannot be combined with the relativity
principles.

We shall not use quantum mechanical principles. Instead we suppose, that the
quantum stochasticity has a geometrical origin. Namely, we suppose, that the
real space-time is described by the geometry of Minkowski only
approximately. This approximation is valid only for large enough values of
 world function for the Minkowski space-time
\begin{equation}
\sigma _M\left( x,x^{\prime }\right) =\sigma _M\left( t,{\bf x},t^{\prime },%
{\bf x}^{\prime }\right) =\frac 12\left( c^2\left( t-t^{\prime }\right)
^2-\left( {\bf x-x}^{\prime }\right) ^2\right)  \label{c3.3}
\end{equation}
where $x=\left\{ t,{\bf x}\right\} $ and $x^{\prime }=\left\{ t^{\prime },%
{\bf x}^{\prime }\right\} $ are coordinates of two points in the space-time,
and $c$ is the speed of light. The world function $\sigma $ is a way of
geometry description \cite{S60}. Recently it was shown that the world
function describes geometry completely \cite{R90,R00}, and any change of
world function leads to a change of geometry and vice versa. Geometry of the
real space-time is described by the world function

\begin{equation}
\sigma \left( x,x^{\prime }\right) =\sigma _M\left( x,x^{\prime }\right)
+D\left( \sigma _M\left( x,x^{\prime }\right) \right)  \label{c3.5}
\end{equation}
Here $\sigma _M$ is the Minkowski world function (\ref{c3.3}), $D$ is a
distortion function
\begin{equation}
D=\left\{
\begin{array}{ccc}
d & {\rm if} & \sigma _M>\sigma _0 \\
0 & {\rm if} & \sigma _M\leq 0
\end{array}
\right. \qquad d=\frac \hbar {2bc}={\rm const},\qquad b\approx 10^{-17}{\rm %
g/cm}  \label{c3.6}
\end{equation}
where $\hbar $ is the quantum constant, and $b$ is a new universal
constant.\ $\sigma _0$ is a constant of the order $d\approx 10^{-21}$cm$^2$.
Correction to the Minkowski world function is small for large values of $%
\sigma _M$, because the characteristic length $\sqrt{d}\approx 10^{-11}$cm
is essential only in the microcosm. Geometry, described by the world
function (\ref{c3.5}), is non-Riemannian. (It is called tubular geometry, or
briefly T-geometry). In T-geometry a motion of free particles is stochastic,
although the world function (\ref{c3.5}) and T-geometry in itself is not
stochastic\footnote{%
It seems rather evident that the free particle motion is stochstic in the
space-time with stochastic geometry \cite{M51,B70,B71}, but a stochastic
motion of a free particle in the deterministic space-time looks rather
unexpected.}. The stochasticity in T-geometry, described by the world
function (\ref{c3.5}), depends on the particle mass. The stochasticity is
large for particles of small masses. It is negligible for particles of
macroscopic mass.

Construction of a geometry on the basis of only world function is a new
conception of geometry. (See for details Ref. \cite{R90,R91,R1995,R00}). Now
from this conception we need only existence of a geometrical stochasticity.
We identify the geometrical stochasticity with the quantum stochasticity and
try to construct a statistical description of microparticles on the basis of
Hamiltonian variational principle. Such an approach has many points in its
favour. First, neither quantum principles, nor new principles are used. All
results are logical corollaries of supposition about the form of the world
function (\ref{c3.5}) of the space-time. Second, such an approach is
relativistic from the outset, and one does not need to combine relativity
principles with quantum mechanical ones and to solve problems, connected
with this integration. Third, the wave function and spin appear to be
attributes of statistical description (not fundamental objects, whose
meaning is obscure). Fourth, interpretation of quantum mechanics appears to
be single-valued. Different versions of quantum mechanics interpretation,
which take place in the axiomatic quantum mechanics, are impossible under
such an approach. In particular, there is a clear distinction between the
individual stochastic system ${\cal S}_{{\rm st}}$ and statistically
averaged system $\left\langle {\cal S}_{{\rm st}}\right\rangle .$ (The last
is the statistical ensemble ${\cal E}_{{\rm p}}\left[ {\cal S}_{{\rm st}%
}\right] $ normalized to one system). In quantum mechanics the same term is
used for both ${\cal S}_{{\rm st}}$ and $\left\langle {\cal S}_{{\rm st}%
}\right\rangle $. It leads to misunderstanding and paradoxes such as paradox
of the Schr\"odinger cat, or EPR paradox. Fifth, a use of the model approach
(instead of axiomatic one) admits one to determine the domain, where the
quantum principles are valid, and quantum mechanics works correctly.

Let now ${\cal S}_{{\rm st}}$ be an electron or some other microparticle.
The system ${\cal S}_{{\rm st}}$ is stochastic, but statistical ensemble $%
{\cal E}_{{\rm p}}\left[ {\cal S}_{{\rm st}}\right] $ of electrons ${\cal S}%
_{{\rm st}}$ is a conservative deterministic dynamic system, and some
variational principle takes place for ${\cal E}_{{\rm p}}\left[ {\cal S}_{%
{\rm st}}\right] $. We suppose that in the zero approximation ${\cal E}_{%
{\rm p}}\left[ {\cal S}_{{\rm st}}\right] $ is described as a statistical
ensemble ${\cal E}_{{\rm p}}\left[ {\cal S}_{{\rm d}}\right] $, where ${\cal %
S}_{{\rm d}}$ is a deterministic relativistic particle, described by the
Hamiltonian
\begin{equation}
H\left( x,{\bf p}\right) =\sqrt{m^2c^4+{\bf p}^2c^2}  \label{b3.1}
\end{equation}
Then variational principle (\ref{a2.24}) for dynamic system ${\cal E}_{{\rm p%
}}\left[ {\cal S}_{{\rm d}}\right] $ has the form
\begin{equation}
{\cal A}_E[\rho ,\varphi ,\bxi ]=\int \rho \{-\sqrt{m^2c^4+{\bf p}^2c^2}%
-b_0[\partial _0\varphi +g^\alpha (\bxi )\partial _0\xi _\alpha ]\}{\rm d}%
^4x,  \label{b3.2}
\end{equation}
where ${\bf p}$ is determined by (\ref{a2.14}), $n=3$. To obtain variational
principle for ${\cal E}_{{\rm p}}\left[ {\cal S}_{{\rm st}}\right] $, one
should take into account stochastic component of the particle momentum. The
mean value ${\bf p}_{{\rm st}}$ of the stochastic component has the form
\begin{equation}
{\bf p}_{{\rm st}}=\hbar {\bf \nabla }\ln \rho  \label{b3.3}
\end{equation}
where $\hbar $ is the Planck quantum constant. In other words, the mean
value ${\bf p}_{{\rm st}}$ of the stochastic component depends on the
collective variable $\rho $ which describes the state of the statistical
ensemble.\ As far as the stochastic components and regular ones are supposed
to be independent, a sum of squares of these components is to be used in the
modified Hamiltonian.

The action for the dynamic system ${\cal E}_{{\rm p}}\left[ {\cal S}_{{\rm st%
}}\right] $ has the form
\begin{equation}
{\cal A}_E[\rho ,\varphi ,\bxi ]=\int \rho \{-\sqrt{m^2c^4+{\bf p}%
^2c^2+\hbar ^2c^2\left( {\bf \nabla }\ln \rho \right) ^2}-b_0[\partial
_0\varphi +g^\alpha (\bxi )\partial _0\xi _\alpha ]\}{\rm d}^4x,
\label{b3.4}
\end{equation}
where ${\bf p}$ is determined by (\ref{a2.14}), $n=3$. This fact may be
interpreted also in the sense that the mass of the electron is modified
\begin{equation}
m^2\rightarrow m_{{\rm q}}^2=m^2+\frac{\hbar ^2}{c^2}\left( {\bf \nabla }\ln
\rho \right) ^2  \label{b3.5}
\end{equation}
In fact, the supposition (\ref{b3.5}) is an origin of supposition (\ref{b3.3}%
). One supposes that consideration of geometric stochasticity leads only to
a change of parameters of the system ${\cal S}_{{\rm d}}$ in ${\cal E}_{{\rm %
p}}\left[ {\cal S}_{{\rm d}}\right] $. These parameters start to depend on
the state density $\rho $ of the ensemble. The system ${\cal S}_{{\rm d}}$ (%
\ref{b3.1}) has only one parameter -- a mass $m$. The dependence (\ref{b3.5}%
) is obtained as a result of averaging over world lines of stochastic
particles \cite{R91}. It is to be invariant with respect to a change of
number of the ensemble elements, i.e. with respect to transformation
\begin{equation}
\rho \rightarrow a\rho ,\qquad a=\mbox{const}>0.  \label{b3.5a}
\end{equation}

Note that the action (\ref{b3.4}) cannot be considered to be an action for
the statistical ensemble of any deterministic dynamic systems ${\cal S}_{%
{\rm d}}$, because the effective Hamiltonian
\begin{equation}
H_{{\rm eff}}=\sqrt{m^2c^4+{\bf p}^2c^2+\hbar ^2c^2\left( {\bf \nabla }\ln
\rho \right) ^2},  \label{b3.6}
\end{equation}
which enters in the Hamilton variational principle, depends now on the
collective variable $\rho $ describing the state of the whole ensemble,
whereas Hamiltonian for any statistical ensemble ${\cal E}_{{\rm p}}\left[
{\cal S}_{{\rm d}}\right] $ of deterministic dynamic systems ${\cal S}_{{\rm %
d}}$ has to depend only on variables of the dynamic system ${\cal S}_{{\rm d}%
}$, as it follows from (\ref{a2.24}). The action (\ref{b3.4}) is an action
for a set of identical deterministic dynamic systems ${\cal S}_{{\rm d}},$
interacting between themselves. Hence, it is not a statistical ensemble of $%
{\cal S}_{{\rm d}}$. In the same time the action (\ref{b3.4}) may be
considered to be an action for a statistical ensemble, because it has the
main property of a statistical ensemble: not to depend on the number of
elements of the statistical ensemble. Mathematically it means that if $\rho $
is substituted by $a\rho $,\ \ $a=$const, ${\cal A}_E[\rho ,\varphi ,\bxi ]$
is substituted by $a{\cal A}_E[\rho ,\varphi ,\bxi ],$ or
\begin{equation}
{\cal A}_E[a\rho ,\varphi ,\bxi ]=a{\cal A}_E[\rho ,\varphi ,\bxi ]
\label{b3.7}
\end{equation}

It is easy to see that the action (\ref{b3.4}) has the property (\ref{b3.7}%
). Hence, the action (\ref{b3.4}) is an action of a statistical ensemble,
but this ensemble cannot be a statistical ensemble of deterministic dynamic
systems, because its elements interact between themselves and are not
independent. It means that the statistical ensemble (\ref{b3.4}) is a
statistical ensemble of nondeterministic dynamic systems. In other words, a
statistical ensemble of nondeterministic systems can be considered to be a
set of interacting deterministic systems, i.e. a stochasticity of dynamic
systems is substituted by interaction of deterministic dynamic systems. Form
of this interaction depends on the form of stochasticity. Considering
different forms of interaction, satisfying the constraint (\ref{b3.7}), one
can label and investigate different forms of stochasticity.

In fact we have no other way of the stochasticity description except for a
reduction of the statistical ensemble ${\cal E}_{{\rm p}}\left[ {\cal S}_{%
{\rm st}}\right] $ to a set ${\cal S}_{{\rm red}}\left[ {\cal S}_{{\rm d}%
}\right] $ of interacting deterministic dynamic systems ${\cal S}_{{\rm d}}$%
.
\begin{equation}
{\cal E}_{{\rm p}}\left[ {\cal S}_{{\rm st}}\right] ={\cal S}_{{\rm red}%
}\left[ {\cal S}_{{\rm d}}\right]  \label{b3.8}
\end{equation}
Character of stochasticity is described by the way of interaction between
deterministic dynamic systems ${\cal S}_{{\rm d}}$ in the set ${\cal S}_{%
{\rm red}}\left[ {\cal S}_{{\rm d}}\right] $ which is a deterministic
dynamic system, consisting of interacting deterministic systems ${\cal S}_{%
{\rm d}}$. We shall refer to ${\cal S}_{{\rm red}}\left[ {\cal S}_{{\rm d}%
}\right] $ as a reduced statistical ensemble, consisting of elements ${\cal S%
}_{{\rm d}}$, although the set ${\cal S}_{{\rm red}}\left[ {\cal S}_{{\rm d}%
}\right] $ of ${\cal S}_{{\rm d}}$ is not a statistical ensemble of ${\cal S}%
_{{\rm d}}$ at all.

Thus, we know the only way of effective mathematical description and
investigation of stochastic dynamic systems. This is a substitution of
stochasticity by an interaction, i.e. reduction of the statistical ensemble $%
{\cal E}_{{\rm p}}\left[ {\cal S}_{{\rm st}}\right] $ to a set ${\cal S}_{%
{\rm red}}\left[ {\cal S}_{{\rm d}}\right] .$ One can describe properties of
stochastic system ${\cal S}_{{\rm st}},$ only referring to the properties of
the reduced statistical ensemble ${\cal S}_{{\rm red}}\left[ {\cal S}_{{\rm d%
}}\right] $. Different kinds of stochasticity are described by a
consideration of different types of interaction between ${\cal S}_{{\rm d}}$
in ${\cal S}_{{\rm red}}\left[ {\cal S}_{{\rm d}}\right] $. We shall show
that the action (\ref{b3.4}) describes a dynamic system which in the
non-relativistic approximation is described by the Schr\"odinger equation.
But there are another reduced statistical ensembles ${\cal S}_{{\rm red}%
}\left[ {\cal S}_{{\rm d}}\right] $ which have the Schr\"odinger equation as
a dynamic equation.

For instance, the reduced statistical ensemble ${\cal S}_{{\rm red}}\left[
{\cal S}_{{\rm d}}\right] $, described by the action
\begin{equation}
{\cal A}_E[\rho ,\varphi ,{\bxi ,}\kappa ]=\int \rho \{-\sqrt{m^2c^4+\hbar
^2c^2\left( \partial _l\kappa ^l+\kappa ^l\kappa _l\right) }-b_0[\partial
_0\varphi +g^\alpha (\bxi )\partial _0\xi _\alpha ]\}{\rm d}^{n+1}x,
\label{b3.9}
\end{equation}
\begin{equation}
\kappa =\left\{ \kappa _0,\kappa _1,\kappa _2,\kappa _3\right\} ,\qquad
\kappa ^l=g^{lj}\kappa _j,\qquad l=0,1,2,3  \label{b3.10}
\end{equation}
where ${\cal S}_{{\rm d}}$ interact via some relativistic quantum $\kappa $%
-field, is also described (under some conditions) by the Klein-Gordon
equation\cite{R98}. In the non-relativistic approximation this equation
reduces to the Schr\"odinger equation. There are another kinds of a reduced
statistical ensemble ${\cal S}_{{\rm red}}\left[ {\cal S}_{{\rm d}}\right] $
which has the Schr\"odinger equation as a dynamic equation under some
conditions. It is not clear a priori, which of these reduced statistical
ensembles ${\cal S}_{{\rm red}}\left[ {\cal S}_{{\rm d}}\right] $ is true.
This question needs further investigation.

Let us return to the action (\ref{b3.4}) and represent it in terms of $\psi $%
-function. In general, introduction of interaction between ${\cal S}_{{\rm d}%
}$ in the action (\ref{b3.4}) is not relativistically covariant. In the
non-relativistic approximation the action (\ref{b3.4}) has the form
\begin{equation}
{\cal A}_E[\rho ,\varphi ,\bxi ]=\int \rho \{-mc^2-\frac{{\bf p}^2}{2m}-%
\frac{\hbar ^2}{2m}\left( {\bf \nabla }\ln \rho \right) ^2-b_0[\partial
_0\varphi +g^\alpha (\bxi )\partial _0\xi _\alpha ]\}{\rm d}^4x,
\label{b3.11}
\end{equation}
where ${\bf p}$ is determined by the relation (\ref{a2.14}).

In terms of the $\psi $-function the action (\ref{b3.11}) is written in the
form
\begin{eqnarray}
A[\psi ,\psi ^{*}] &=&\int \{\frac{ib_0}2(\psi ^{*}\partial _0\psi -\partial
_0\psi ^{*}\cdot \psi )-mc^2\rho -\frac{\hbar ^2\left( {\bf \nabla }\rho
\right) ^2}{2m\rho }  \nonumber \\
&&+\frac{b_0^2}{8\rho m}(\psi ^{*}{\bf \nabla }\psi -{\bf \nabla }\psi
^{*}\cdot \psi )^2\}{\rm d}^4x,  \label{b3.12}
\end{eqnarray}
where $\rho \equiv \psi ^{*}\psi $.

Let the function $\psi $ have $k$ components. Regrouping components of the
function $\psi $ in the action (\ref{b3.12}), one obtains the action in the
form
\[
{\cal A}_E[\psi ,\psi ^{*}]=\int \{\frac{ib_0}2(\psi ^{*}\partial _0\psi
-\partial _0\psi ^{*}\cdot \psi )-\frac{b_0^2}{2m}{\bf \nabla }\psi
^{*}\cdot {\bf \nabla }\psi
\]
\begin{equation}
+\frac{b_0^2}4\sum\limits_{\alpha ,\beta =1}^kQ_{\alpha \beta ,\gamma
}^{*}Q_{\alpha \beta ,\gamma }\rho +\frac{b_0^2-\hbar }{8\rho m}^2(\nabla
\rho )^2-mc^2\rho \}{\rm d}^4x,\qquad \rho \equiv \psi ^{*}\psi  \label{s5.9}
\end{equation}
where
\begin{equation}
Q_{\alpha \beta ,\gamma }={\frac 1{\psi ^{*}\psi }}\left|
\begin{array}{cc}
\psi _\alpha & \psi _\beta \\
\partial _\gamma \psi _\alpha & \partial _\gamma \psi _\beta
\end{array}
\right| ,\qquad \alpha ,\beta =1,2,\ldots k\qquad \gamma =1,2,3
\label{s5.11}
\end{equation}
and $Q_{\alpha \beta ,\gamma }^{*}$ is complex conjugate to $Q_{\alpha \beta
,\gamma }$.

In the simplest case, when the $\psi $-function has only one component, all
quantities $Q_{11,\gamma }={0},\quad \gamma =1,2,3$, and the motion of the
ensemble particles is irrotational. Then the action (\ref{s5.9}) reduces to
the form
\[
{\cal A}_E[\psi ,\psi ^{*}]=\int \{\frac{ib_0}2(\psi ^{*}\partial _0\psi
-\partial _0\psi ^{*}\cdot \psi )-\frac{b_0^2}{2m}{\bf \nabla }\psi
^{*}\cdot {\bf \nabla }\psi
\]
\begin{equation}
-mc^2\rho +\frac{b_0^2-\hbar }{8\rho m}^2(\nabla \rho )^2\}{\rm d}^4x,\qquad
\rho \equiv \psi ^{*}\psi  \label{b3.13}
\end{equation}

Dynamic equation, generated by the action (\ref{b3.13}) is nonlinear due to
the last term in the action (\ref{b3.13}). Equating the arbitrary
integration constant $b_0$ to $\hbar $ $\left( b_0=\hbar \right) ,$ one
obtains instead of (\ref{b3.13})
\begin{equation}
{\cal A}_E[\psi ,\psi ^{*}]=\int \{\frac{i\hbar }2(\psi ^{*}\partial _0\psi
-\partial _0\psi ^{*}\cdot \psi )-\frac{\hbar ^2}{2m}{\bf \nabla }\psi
^{*}\cdot {\bf \nabla }\psi -mc^2\psi ^{*}\cdot \psi \}{\rm d}^4x
\label{b3.14}
\end{equation}

Note that the equivalence of the actions (\ref{b3.13}) and (\ref{b3.14}) can
be verified directly, making in (\ref{b3.13}) a change of variables
\begin{equation}
\psi \to \tilde \psi =|\psi |\exp \left( \frac \hbar {b_0}\log \frac \psi
{|\psi |}\right) ,  \label{b3.14a}
\end{equation}
which transforms the action (\ref{b3.13}) into the action (\ref{b3.14})

The action (\ref{b3.14}) generates linear dynamic equation for the $\psi $%
-function.

\begin{equation}
{\frac{\delta {\cal A}}{\delta \psi _\alpha ^{*}}}=i\hbar \partial _0\psi +%
\frac{\hbar ^2}{2m}{\bf \nabla }^2\psi -mc^2\psi =0  \label{b3.15}
\end{equation}
After substitution $\psi =\Psi \exp \left( \frac{mc^2t}{i\hbar }\right) $
the dynamic equation (\ref{b3.15}) transforms to the Schr\"odinger equation
for a free particle
\begin{equation}
i\hbar \partial _0\Psi +\frac{\hbar ^2}{2m}{\bf \nabla }^2\Psi =0
\label{b3.16}
\end{equation}
It means that in some cases the $\psi $-function can coincide with the wave
function.

Let us compare actions (\ref{b3.13}) and (\ref{b3.14}). They differ in a
choice of the integration constant $b_0$ and describe the same dynamic
system. The action (\ref{b3.13}) contains only one quantum term (i.e. the
term, containing the quantum constant $\hbar $). This term $-\hbar ^2(\nabla
\rho )^2/8\rho m$ describes the energy density of stochastic component of
motion. Other terms, containing integration constant $b_0$, are usual
dynamical terms. In the action (\ref{b3.14}) practically all terms contain
the quantum constant $\hbar $, and should be interpreted as quantum terms.
Extension of quantum properties to all dynamical terms is the price which is
paid for linearity of dynamic equation (\ref{b3.15}).

Separation of quantum and classical properties is very simple in the action (%
\ref{b3.13}). Setting $\hbar =0$ in (\ref{b3.13}), one suppresses all
quantum properties, and remaining terms describe a pure ensemble of
deterministic classical particles. In the action (\ref{b3.14}) one cannot
set $\hbar =0$, because all dynamic terms contain $\hbar $. Separation of
the classical part of the action in (\ref{b3.14}) is rather complicated
procedure. From viewpoint of statistical description a linearity of dynamic
equation (\ref{b3.15}) is an accidental circumstance, connected with a
special form $-\hbar ^2(\nabla \rho )^2/8\rho m$ of the stochastic component
energy density and with irrotational character of the quantum fluid flow. If
this flow is rotational, and $\psi $-function has more, than one component,
dynamic equations are not linear. One cannot be sure, that completely
relativistic statistical description can reduce to a linear dynamic
equation, because in this case the stochastic component energy density has
the form, which differs from $-\hbar ^2(\nabla \rho )^2/8\rho m$.
Nevertheless there exists such a relativistic statistical description, which
generates a linear dynamic equation for one-component $\psi $-function (the
action (\ref{b3.9}) generates the Klein-Gordon equation for one-component $%
\psi $-function \cite{R98}).

Note that the term $\rho U_{{\rm st}}=\hbar ^2(\nabla \rho )^2/8\rho m$,
describing stochastic component energy density, becomes to be large, if
there is a local increase of density $\rho .$ Let $\delta x$ be the linear
size of the region of the density increase, then $|\nabla \rho |\approx \rho
/\delta x$, and the stochastic component energy $U_{{\rm st}}\approx \frac{%
^{\hbar ^2}}{8m\left( \delta x\right) ^2}$ becomes very large, provided $%
\delta x$ is small enough. After expansion of the quantum fluid the local
density increase disappears and the stochastic component energy $U_{{\rm st}%
} $ transforms to the kinetic energy $p^2/2m$. The particle momentum $p$
becomes to be of the order $p\approx \sqrt{2mU_{{\rm st}}}\approx \frac
\hbar {2\delta x}$. This relation is an origin of the indeterminacy
relation. It means that the well known indeterminacy relations is a
corollary of the way as the stochastic component energy $U_{{\rm st}}$
depends on the statistical ensemble density $\rho $.

Of course, linearity of dynamic equations is very convenient and useful, but
it does not mean that linearity of dynamic equations may be considered to be
a principle for construction of relativistic quantum theory.

\end{document}